\documentclass[aps,prb,twocolumn,showpacs,amssymb]{revtex4-1}
\usepackage{amsmath,graphics,epsfig,mathrsfs}
\usepackage{bm}

\newcommand{\etal}{{\em et al.~}}

\begin{document}

\title{Spin Hall effect versus Rashba torque: a Diffusive Approach}
\author{Aurelien Manchon}
\email{aurelien.manchon@kaust.edu.sa}
\affiliation{King Abdullah University of Science and Technology (KAUST),Physical Science and Engineering, Thuwal 23955-6900, Saudi Arabia.}
\date{\today}
\begin{abstract}
Current-driven magnetization dynamics of single ferromagnets in heavy metal/ferromagnet bilayers has been recently realized. In this work, spin torque induced by so-called Rashba spin-orbit coupling and spin Hall effect are considered within a diffusive model. The dependence of the resulting torque as a function of the thicknesses of the ferromagnet and heavy metal is analyzed. We show that (i) both torques are on the form ${\bf T}=T_{||}{\bf m}\times{\bf y}+T_\bot{\bf m}\times({\bf y}\times{\bf m})$, (ii) the ratio $T_{||}/T_\bot$ strongly depends on the thickness of the layers and (iii) the thickness dependence of the spin torque provides an indication of the origin of the (Rashba- or spin Hall effect-induced) spin torque .
\end{abstract}
\maketitle
\section{Introduction}
The electrical manipulation of magnetic ultrathin layers is presently attracting increasing interest due to the possible technological applications \cite{Slonc96,review1,appli}. Current-driven magnetization switching and excitations have been realized in spin-valves nanopillars and non-local configuration as well as magnetic domain walls \cite{chapter1,chapter2}. Interestingly, recent progress has been achieved towards the manipulation of single homogeneous ferromagnets in the absence of a polarizer or magnetic texture \cite{nano,mihai1, pi, mihai2, mihai3,suzuki,liu,chernyshov,fang,endo}. These new configurations take advantage of spin-orbit coupling to manipulate the magnetization. An interesting example is the voltage-controlled anisotropy \cite{nano}, where the perpendicular magnetic anisotropy is controlled by an external gate voltage applied across a dielectric layer. In this configuration, no current flow is involved and the microscopic origin of the anisotropy change is attributed to voltage control of the band filling \cite{tsymbal}, also modeled by an effective interfacial Rashba spin-orbit coupling \cite{manchon-prb2}.\par


Alternatively, current-driven magnetization switching and excitation in single ferromagnets have been realized in the current-in-plane (CIP) configuration. These structures consist of HM/F/MOx stacks where HM designates a heavy metal (Pt, Ta), F is a transition metal ferromagnet and MOx is a metal oxide (MgO or AlOx) (see Refs. \onlinecite{mihai1, pi, mihai2, mihai3,suzuki,liu}). This effect has also been reported in strained dilute semiconductors \cite{chernyshov,fang,endo}. The observation of current-driven magnetization control in the absence of a spin polarizer has been interpreted along two different schemes. One relies on the Rashba torque previously proposed in Refs. \onlinecite{manchon-prb,others} and recently extended in Refs. \onlinecite{wang-manchon-2011,kim}. This torque, derived in the case of ferromagnetic 2-dimensional electron gas (2DEG) in the presence of Rashba spin-orbit coupling \cite{rashba} is dominated by a field-like term along ${\bf m}\times{\bf y}$ (${\bf y}$ is the in-plane direction transverse to the current injection and ${\bf m}$ is the magnetization direction) and possesses a correction on the form of an (anti-)damping term along ${\bf m}\times({\bf y}\times{\bf m})$. Although applying the Rashba model to realistic trilayer structures involving ultrathin metallic layers remains questionable, Rashba spin-orbit coupling has been observed in a wide variety of metallic interfaces \cite{int}. Therefore, structures with asymmetric interfaces such as HM/F/MOx are appropriate to display interfacial Rashba interactions.\par

Another possible mechanism relies on the Spin Hall effect\cite{she} (SHE) torque proposed by Liu {\em et al.} \cite{liu}. In this case, the strong SHE present in the HM bottom layer injects a spin current normally to the interfaces with a polarization along ${\bf j}\times{\bf z}={\bf y}$, where ${\bf j}$ is the direction of the current injection and ${\bf z}$ is the normal to the interfaces. This SHE torque is equivalent to a torque arising from a polarizing layer that would be located below the F layer with its magnetization along ${\bf y}$ and the current injection along ${\bf z}$. By pursuing the analogy, this torque is dominated by a (anti-)damping torque along ${\bf m}\times({\bf y}\times{\bf m})$ and a field-like torque along ${\bf m}\times{\bf y}$. This field-like torque is usually disregarded in spin-valves due to the short spin dephasing length \cite{zlf,stiles2002}. However, when the layer thickness is comparable to the dephasing length, the spin current is not totally absorbed in the magnetic free layer and one can reasonably expect a field-like component to emerge \cite{zlf}.\par

Consequently, although the two torques, SHE and Rashba-driven, are expected to be dominated by different components (anti-damping torque for SHE and field-like torque for Rashba), they both adopt the same geometrical form ${\bf T}=T_{||}{\bf m}\times{\bf y}+T_\bot{\bf m}\times({\bf y}\times{\bf m})$. A major issue is then to find a way to distinguish between the two origins. In this work, we model the spin transport in a HM/F bilayer stack in three cases. In the first case, only SHE is allowed in the HM layer. In the second case, Rashba spin-orbit coupling is assumed to be present at the top F/MOx interface whereas SHE is absent in the HM layer. In the third case, Rashba is present at the bottom HM/F interface and SHE is neglected in HM. We find that the spin dynamics is different is these three cases leading to different thickness dependencies of the spin torque.

\section{Transport Formalism}

The transport formalism itself is subjected to discussion. The configuration of the samples under investigation are thin multilayers in which a current is injected in the plane (CIP). CIP configurations have been intensively studied both theoretically and experimentally in the early 90's, soon after the discovery of Giant MagnetoResistance \cite{gmr} (GMR). The tremendous success of spin-valves configuration in which the current is injected perpendicular to the plane \cite{sv} (CPP) has partially occulted the important discoveries established in CIP samples. From the theory viewpoint, whereas the semi-classical diffusion formalism developed by Valet and Fert \cite{vf} (VF) has been tremendously successful in describing CPP spin-valves, it simply does not apply in the case of CIP multilayers. One striking difference is that while spin transport is governed by the spin diffusion length $\lambda_{sf}$ in CPP spin-valves, it is characterized by the spin-dependent mean free path $\lambda_e^{\sigma}$ in CIP multilayers. An instructive consequence of this seminal difference is that no CIP-GMR effect can be captured when applying VF approach to CIP systems. The reason is that Bolztmann formalism, such as the one proposed by Camley and Barnas \cite{cb} (CB), treats the evolution of the spin-dependent carrier distribution function $g^\sigma(t,{\bf r}, {\bf v})$, which depends on the spin $\sigma$, time $t$, position ${\bf r}$ and particle velocity ${\bf v}$. In contrast, VF theory tracks the behavior of semi-classical quantities such as spin accumulation and spin current which implies averaging the carrier distribution over the velocity ${\bf v}$. Therefore, whereas CB formalism explicitly accounts for the particle velocity ${\bf v}$, it disappears in VF theory (for details, see Ref. \onlinecite{vf}).\par

\begin{figure}
	\centering
		\includegraphics[width=8cm]{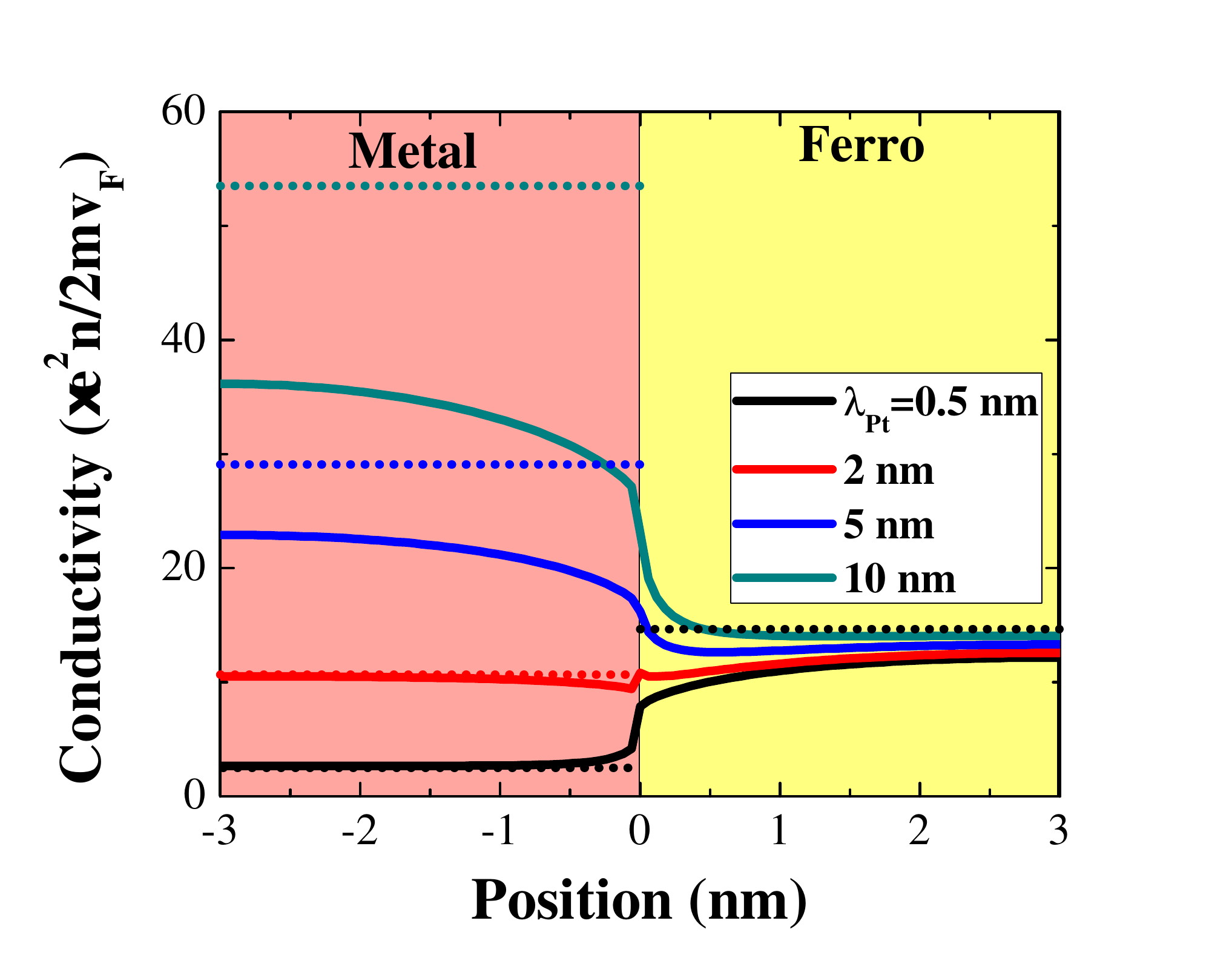}
	\caption{\label{fig:Fig0}(Color online) Distribution of the conductivity in the direction normal to the plane of the bilayer. The dotted lines are the expected values in the diffusive approximation. The parameters are the same as in Ref. \onlinecite{cb}, $\lambda_F^\uparrow$= 5 nm and $\lambda_F^\downarrow$= 0.5 nm.}
\end{figure}

To demonstrate the added value of Boltzmann formulation of the spin transport, we model a normal metal/ferromagnet bilayer in CIP configuration following Camley and Barnas \cite{cb}. In this approach, the semiclassical non-equilibrium electron distribution $g^\sigma(t,{\bf r}, {\bf v})$ is governed by Boltzmann transport equation
\begin{equation}\label{eq:bol}
\frac{dg^\sigma}{dt}=\frac{\partial g^\sigma}{\partial t}+{\bf v}\cdot{\bm\nabla}_{\bf r}g^\sigma+{\dot{\bf v}}\cdot{\bm\nabla}_{\bf v}g^\sigma=-\frac{g^\sigma}{\tau_\sigma},
\end{equation}
where $\tau_\sigma$ is the spin-dependent momentum scattering time. In a bilayer homogeneous along the ($x$-$y$) plane and perpendicular to the $z$ direction, the solution of Eq. (\ref{eq:bol}) is on the form $g^\sigma_\pm=\frac{eE\tau^\sigma}{m}(1+Ae^{\pm z/\tau_\sigma|v_z|})$. This approach has the advantage to explicitly account for the momentum distribution. In order to illustrate its implication on CIP transport, we solve Eq. (\ref{eq:bol}) following Ref. \onlinecite{cb}, assuming diffusive interfaces and totally reflective outer boundaries. The results are displayed in Fig. \ref{fig:Fig0}, together with the expected values in the diffusive limit, $\sigma=n e^2\tau/m$ (dotted lines). Interestingly, where the electron mean free path $\lambda_e=v_F\tau$ is much smaller than the layers thickness ($\lambda_e/d\ll1$), the conductivity is very similar to the one expected in the diffusive limit. One the other hand, when the mean free path becomes on the same order as or larger than the layers thickness ($\lambda_e/d\gg1$), the conductivity obtained within CB approach strongly differs from the diffusive limit.\par

We retain two important aspects from the results in Fig. \ref{fig:Fig0}: (i) the current distribution is not homogeneous inside the layers and (ii) the effective conductance of the ultrathin layers is likely to be very different from the bulk conductances. Nevertheless, in realistic ultrathin layers the grains size and interfaces roughness are expected to play an important role and to reduce the effective mean free path. Therefore, item (ii) is not critical since the actual conductivity can be captured through an effective parameter.\par

More importantly, the smooth variation of the current distribution at the vicinity of the interface may have quantitative consequences on the actual spin dynamics at the interface (where spin dephasing takes place \cite{stiles2002}) and then on the magnitude of injected spin current. However, it can be argued that the conductivity jump at the interface remains quite sharp and therefore, the added value of Boltzmann formalism to accurately treat spin transfer torque in such bilayers may actually be limited.\par

\begin{figure}
	\centering
		\includegraphics[width=10cm]{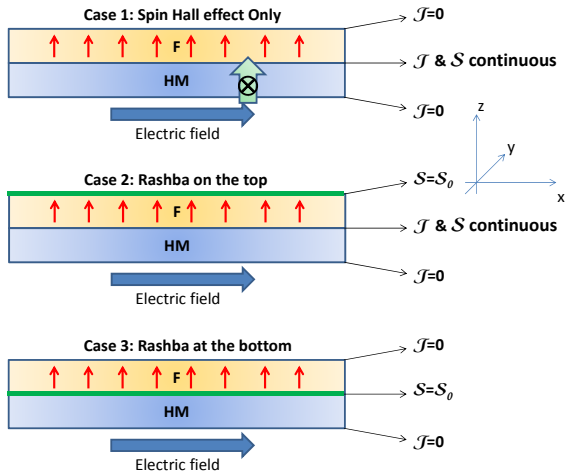}
	\caption{\label{fig:Fig1}(Color online) Schematics of the bilayer structure and boundary conditions for the three cases considered.}
\end{figure}

As a consequence, in the remainder of the paper we use a semi-classical diffusive model in line with VF theory. In general the spin and charge currents are written \cite{vedy}
\begin{eqnarray}\label{eq:1}
{\cal J}&=&-{\cal D}{\bm \nabla}\otimes{\bf S}-\alpha_H{\cal D}\left[{\bm \mu}\times{\bm\nabla}n\right]\otimes{\bm \mu}\\
{\bf j}_e&=&-{\cal D}{\bm \nabla}n-\beta {\cal D}{\bm \nabla}({\bf S}\cdot{\bf m})+a_0^3\alpha_H{\cal D}{\bf S}\times{\bm\nabla}n
\end{eqnarray}
where ${\cal J}$ is the spin current density, ${\bf S}$ is the itinerant spin density (spin accumulation), $n$ is the charge accumulation, ${\cal D}$ is the diffusion coefficient, $\alpha_H$ is the Hall angle, $\beta$ is the bulk spin asymmetry parameter and $a_0^3$ is the volume of the unit cell. The unit vector ${\bm\mu}$ is the direction of the spin projection. The spin accumulation dynamics is given by
\begin{eqnarray}\label{eq:2}
&&\frac{\partial {\bf S}}{\partial t}=-{\bm\nabla}\cdot{\cal J}-\frac{1}{\tau_J}{\bf S}\times{\bf m}-\frac{1}{\tau_{\phi}}{\bf m}\times({\bf S}\times{\bf m})-\frac{{\bf S}}{\tau_{sf}}
\end{eqnarray}
where $\tau_J$ is the spin precession time, $\tau_\phi$ is the spin dephasing length and $\tau_{sf}$ is the spin diffusion length. The system Eqs. (\ref{eq:1})-(\ref{eq:2}) is solved assuming the following boundary conditions, illustrated in Fig. \ref{fig:Fig1}:
\begin{itemize}
\item SHE only: total current reflection (${\cal J}$=0) at both outer boundaries and continuity of spin current and spin accumulation at the HM/F interface.
\item Rashba at F/MOx interface: total current reflection (${\cal J}$=0) at the outer HM boundary, spin accumulation determined by Rashba-induced spin density ${\bf S}_0$ at the F/MOx boundary and continuity of spin current and spin accumulation at the HM/F interface.
\item Rashba at HM/F interface: total current reflection (${\cal J}$=0) at both outer boundaries and spin accumulation determined by Rashba-induced spin density ${\bf S}_0$ at the HM/F interface.
\end{itemize}
We do not provide the detail of the derivation here. The spin accumulation is directly extracted from Eqs. (\ref{eq:1})-(\ref{eq:2}) and the spin transfer torque is defined as
\begin{eqnarray}\label{eq:st}
&&{\bf T}=\frac{1}{\Omega}\int_{\Omega}d\Omega\left(-{\bm \nabla}\cdot{\cal J}-\frac{1}{\tau_{sf}}{\bf m}\right),
\end{eqnarray} namely the spatial change of spin current, compensated by the spin relaxation term. $\Omega$ is the volume of the magnetic layer. In the remaining of the article, we consider the magnetization oriented along $z$, i.e. out-of-plane.

\section{Results}
\subsection{General remarks}
Before entering in the details of the numerics, let us first comment on the general form of the spin transfer torque. The original concept of spin transfer torque proposed by Slonczewski relied on the transfer of transverse spin current to the local magnetization \cite{Slonc96}, ${\bm T}=-{\bm\nabla}\cdot{\cal J}$. This definition describes the core of the phenomenon, but in realistic structures, spin diffusion may modify the form of the spin torque. By exploiting the definition given in Eq. (\ref{eq:st}) and Eq. (\ref{eq:2}), one can expression the resulting local torque in terms of spin current divergency
\begin{eqnarray}
&&{\bm T}=-{\bm\nabla}\cdot{\cal J}+\frac{\beta}{1+\xi^2}{\bm m}\times{\bm\nabla}\cdot{\cal J},\label{eq:torqueJ}\\
&&\beta=\frac{\tau_J}{\tau_{sf}},\;\xi=\tau_J\left(\frac{1}{\tau_{sf}}+\frac{1}{\tau_\phi}\right).\nonumber
\end{eqnarray}
Equation (\ref{eq:torqueJ}) indicates that the spin torque is in general not simply given by the divergency of the spin current but possesses an additional component ${\bf m}\times{\bm\nabla}\cdot{\cal J}$ arising from the presence of spin diffusion. Therefore, there is always a correction to the main torque component when spin relaxation is present.\par

In the following sections, we address the dependence of the torques as a function of the layer thicknesses. The magnitude of the SHE torque is proportional to the current density in the heavy normal metal, $j_{HM}$, whereas the magnitude of the Rashba torque is proportional to the interfacial current $j_i$. In order to keep the contribution of the spin dynamics explicit, we assume that $j_{HM}$ and $j_i$ are kept constant when varying the layers thicknesses. In a realistic experiment, it is much easier to apply the same total current density. Assuming a simple circuit description of the bilayer, one can relate these current densities to the total injected current density $j_T$
\begin{eqnarray}\label{eq:jn}
&&j_{HM}=\frac{\sigma_{HM} d_{HM}}{\sigma_{HM} d_{HM}+\sigma_Fd_F}j_T,\\\label{eq:ji}
&&j_i=\frac{\sigma_it_i}{\sigma_it_i+\sigma_{HM} d_{HM}+\sigma_Fd_F}j_T,
\end{eqnarray}
where $\sigma_\nu$ and $d_\nu$ ($\nu=i,HM,F$) are the conductivity and (effective) thickness of the interface, normal metal and ferromagnet, respectively. This renormalization produces an additional thickness dependence that must be taken into account to properly describe the thickness dependence of the torque. Furthermore, in the light of the discussion about CB model of CIP transport, we also need to keep in mind that the effective conductivity of an ultrathin layer depends on the parameter $\lambda_e/d$, which implies that the conductivities $\sigma_\nu$ are also thickness dependent. To avoid unnecessary complexity in the analysis of the results, we focus on the influence of the layers thickness on the spin dynamics itself ($j_{N,i}$ is kept constant).

\subsection{Case 1: Spin Hall effect Torque}

\begin{figure}
	\centering
		\includegraphics[width=8cm]{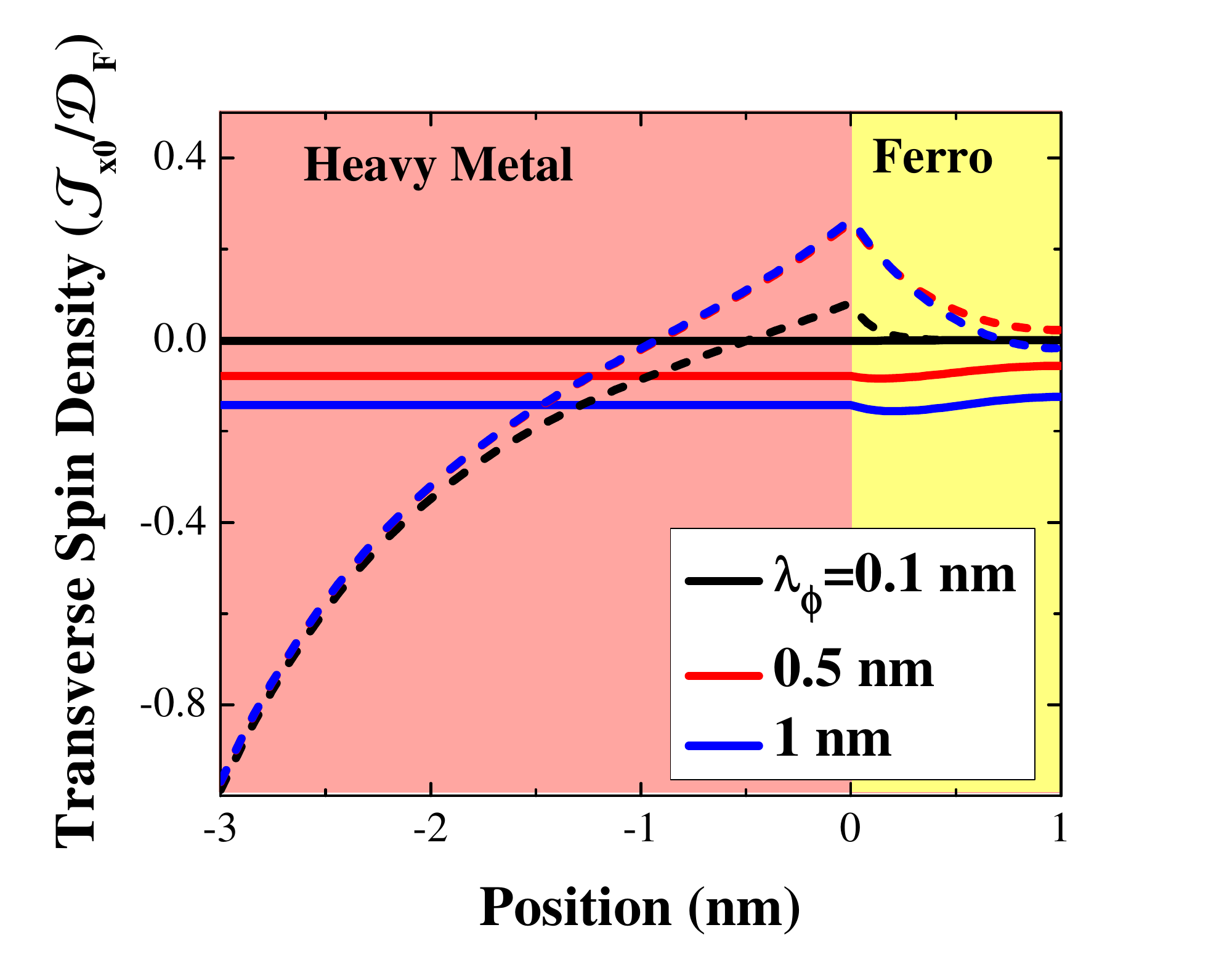}
	\caption{\label{fig:Fig2}(Color online) Case 1: Transverse spin accumulation $S_x$ (solid lines) and $S_y$ (dashed lines) profile along $z$ in the presence of SHE in the heavy metal only, for different spin dephasing lengths. The other parameters are $\lambda_{sf}^F$=15 nm, $\lambda_J^F$=0.5 nm, $\lambda_{sf}^{HM}$=1 nm, $d_F$= 1 nm, $d_{HM}$=3 nm and ${\cal D}_F/{\cal D}_{HM}=1$.}
\end{figure}

In this case, a spin Hall current polarized along ${\bf y}$ is generated in the HM layer and impinges onto the F layer. Fig. \ref{fig:Fig2} displays the spin accumulation profile $S_x$ and $S_y$ are a function of the position in the bilayer for different values of the spin dephasing. Following the definition of the spin torque in Eq. (\ref{eq:st}), the component $S_x$ produces the (anti-)damping torque and the component along $S_y$ produces the field-like torque. As expected, the $S_y$ component vanishes for very short spin dephasing. In this case, the transverse spin current ${\cal J}=\alpha_H{\bf j}\times{\bf y}\otimes{\bf y}$ is fully absorbed by at the HM/F interface and the torque reduces to the pure (anti-)damping torque.\par

\begin{figure}
	\centering
		\includegraphics[width=8cm]{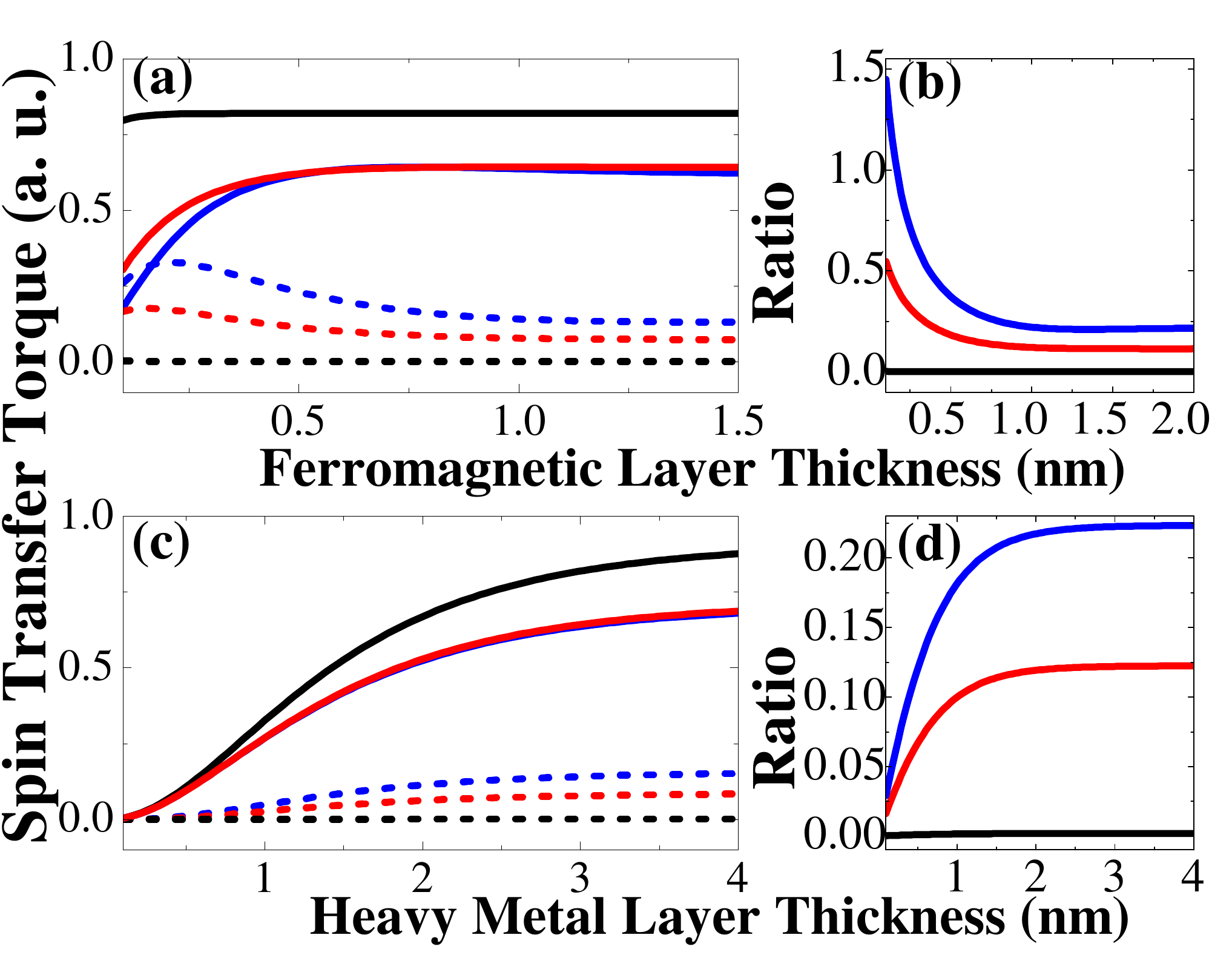}
	\caption{\label{fig:Fig3}(Color online) Case 1: In-plane ($T_{||}$ - solid lines) and out-of-plane torques ($T_\bot$ - dashed lines) as a function of the F (a) and HM (c) layers thickness for different spin dephasing lengths $\lambda_\phi$; spin torque ratio $T_\bot/T_{||}$ as a function of the F (b) and HM (d) layers thickness. The parameters are the same as in Fig. \ref{fig:Fig2}.}
\end{figure}

The thickness dependence of the resulting absolute torque $d {\bf T}$ is represented in Fig. \ref{fig:Fig3}. Note that the torque is multiplied by the distance in order to the remove the $1/d$-dependence that naturally appears after volume averaging. As mentioned in the introduction, in the case of spin Hall effect, the in-plane torque ($\propto{\bf m}\times({\bf y}\times{\bf m})$) dominates over the out-of-plane torque ($\propto{\bf m}\times{\bf y}$). Whereas the absolute torque decreases for very thin F layers ($\lambda_\phi/d_F\geq1$), it remains essentially unaffected by the Co thickness for $d_F>\lambda_\phi$.\par

Conversely, since the spin current is generated by asymmetric spin scattering in the bulk of the HM layer, increasing the HM layer thickness increases the spin torque in the range $\lambda_{sf}^{HM}>d_{HM}>0$. When the thickness of the heavy metal exceeds the spin diffusion length ($d_{HM}>\lambda_{sf}^{HM}$), the amount of spin Hall current injected into the ferromagnet saturates due to spin-flip scattering (the additional HM thickness is inefficient in creating more spin Hall current). The thickness dependence is on the form $\propto1-\cosh^{-1}d_{HM}/\lambda_{sf}^{HM}$, as proposed by Liu {\em et al.} \cite{liu}.
A last remark is that the ratio $T_{\bot}/T_{||}$ is also thickness dependent, as shown in Fig. \ref{fig:Fig3}(b) and (d). While the ratio saturates towards the bulk value given by Eq. (\ref{eq:torqueJ}), it increases when decreasing the thickness of the ferromagnetic layer and decreases when decreasing the thickness of the heavy metal. Note however that a significant enhancement of the perpendicular component is only reached for extremely thin layers. Therefore, one can reasonably expect that the SHE torque reduces essentially to an in-plane torque ($\propto{\bf m}\times({\bf y}\times{\bf m})$) with a measurable dependence as a function of the HM layer thickness.

\subsection{Case 2: Rashba on the F surface}
\begin{figure}
	\centering
		\includegraphics[width=8cm]{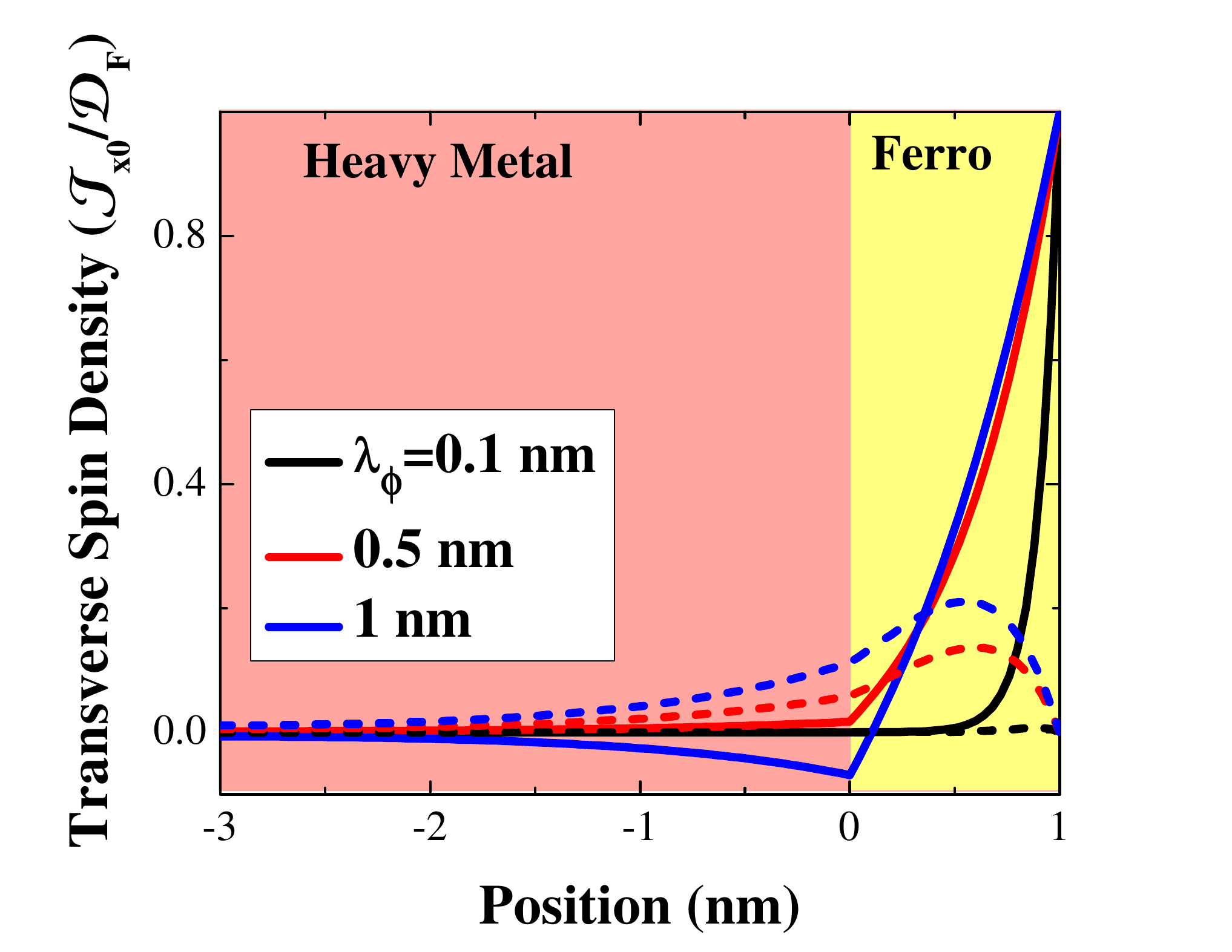}
	\caption{\label{fig:Fig4}(Color online) Case 2: Transverse spin accumulation $S_x$ (solid lines) and $S_y$ (dashed lines) profile along $z$ in the presence of Rashba spin-orbit coupling at the F/MOx interface only, for different spin dephasing lengths. The parameters are the same as in Fig. \ref{fig:Fig2}.}
\end{figure}

In this case, the interfacial Rashba spin-orbit coupling is expected to produce current-induced non-equilibrium spin density along ${\bf S}_0=S^0_x{\bf x}+S^0_y{\bf y}$ \cite{wang-manchon-2011}. For simplicity, we consider that the Rashba torque produces only $S_x^0$ non-zero component ($S_x^0>>S_y^0\approx$0). As shown in Fig. \ref{fig:Fig4}, this interfacial spin accumulation decays away from the surface producing a complex dynamics that generates both $S_x$ and $S_y$ spin density components in the ferromagnet. For a vanishing spin dephasing length, only $S_x$ component survives which produces a field-like Rashba torque \cite{manchon-prb}.\par
\begin{figure}
	\centering
		\includegraphics[width=8cm]{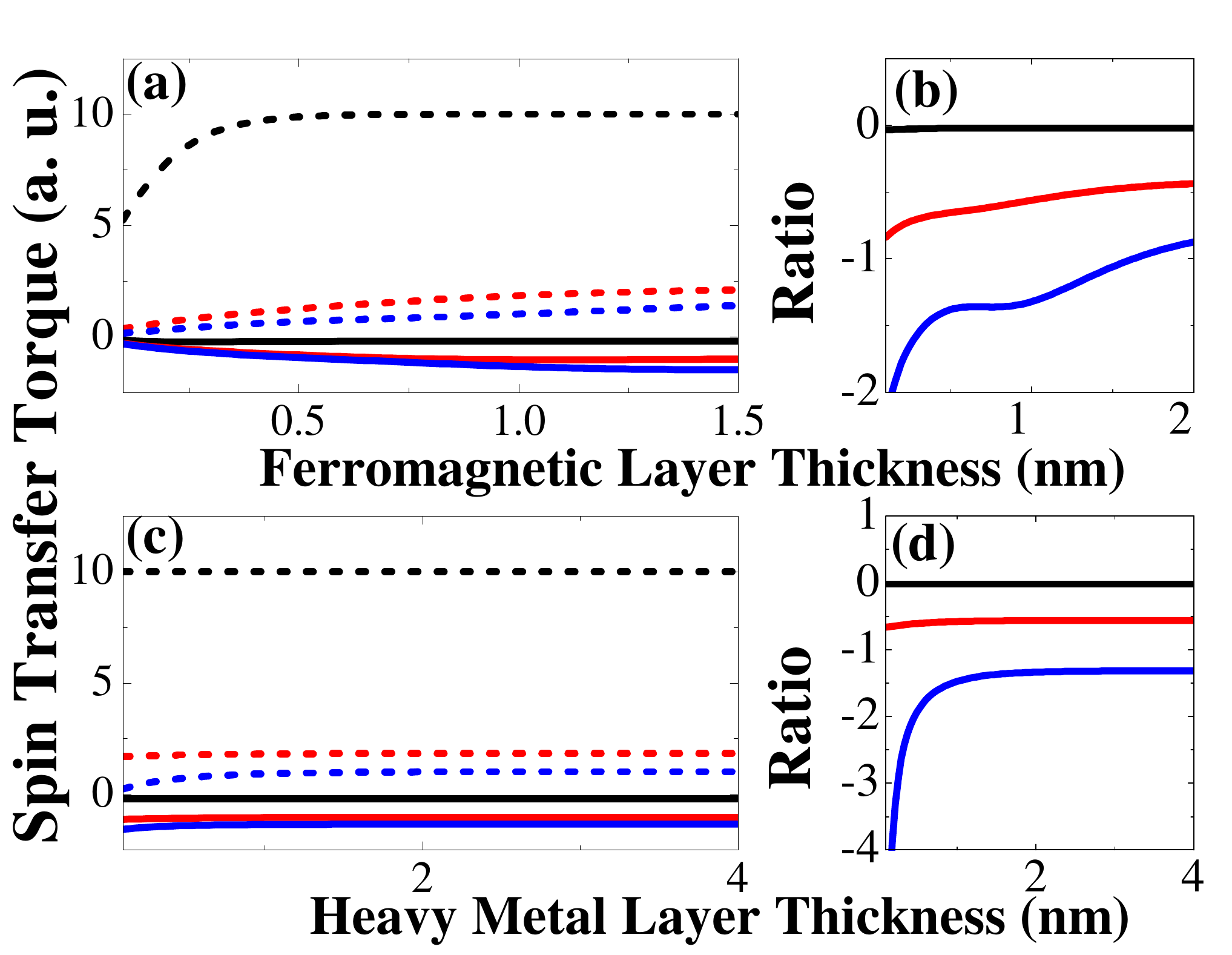}
	\caption{\label{fig:Fig5}(Color online) Case 2: In-plane ($T_{||}$ - solid lines) and out-of-plane torques ($T_\bot$ - dashed lines) as a function of the F (a) and HM (c) layers thickness for different spin dephasing lengths $\lambda_\phi$; spin torque ratio $T_{||}/T_\bot$ as a function of the F (b) and HM (d) layers thickness. The parameters are the same as in Fig. \ref{fig:Fig2}.}
\end{figure}

The thickness dependence of the resulting absolute torque $d {\bf T}$ is represented in Fig. \ref{fig:Fig5}. Whereas no significant thickness dependence is observed when varying the heavy metal thickness (Fig. \ref{fig:Fig5}(c)), a sizable thickness dependence is observed as a function of the thickness of the ferromagnet (Fig. \ref{fig:Fig5}(a)). Interestingly, we note that the in-plane torque can be significantly larger than the bulk value in the case of ultrathin ferromagnetic layers, as shown by the ratio $T_{||}/T_\bot$ in Fig. \ref{fig:Fig5}(b). This additional in-plane component has been theoretically shown to have a significant impact on the current-driven domain wall motion in the presence of Rashba torque \cite{kim}. Therefore, while no thickness dependence is expected from the HM layer, tuning the F layer thickness may strongly enhance the Rashba-induced in-plane torque and have dramatic effects on current-driven magnetization dynamics.

\subsection{Case 3: Rashba HM/F interface}
\begin{figure}
	\centering
		\includegraphics[width=8cm]{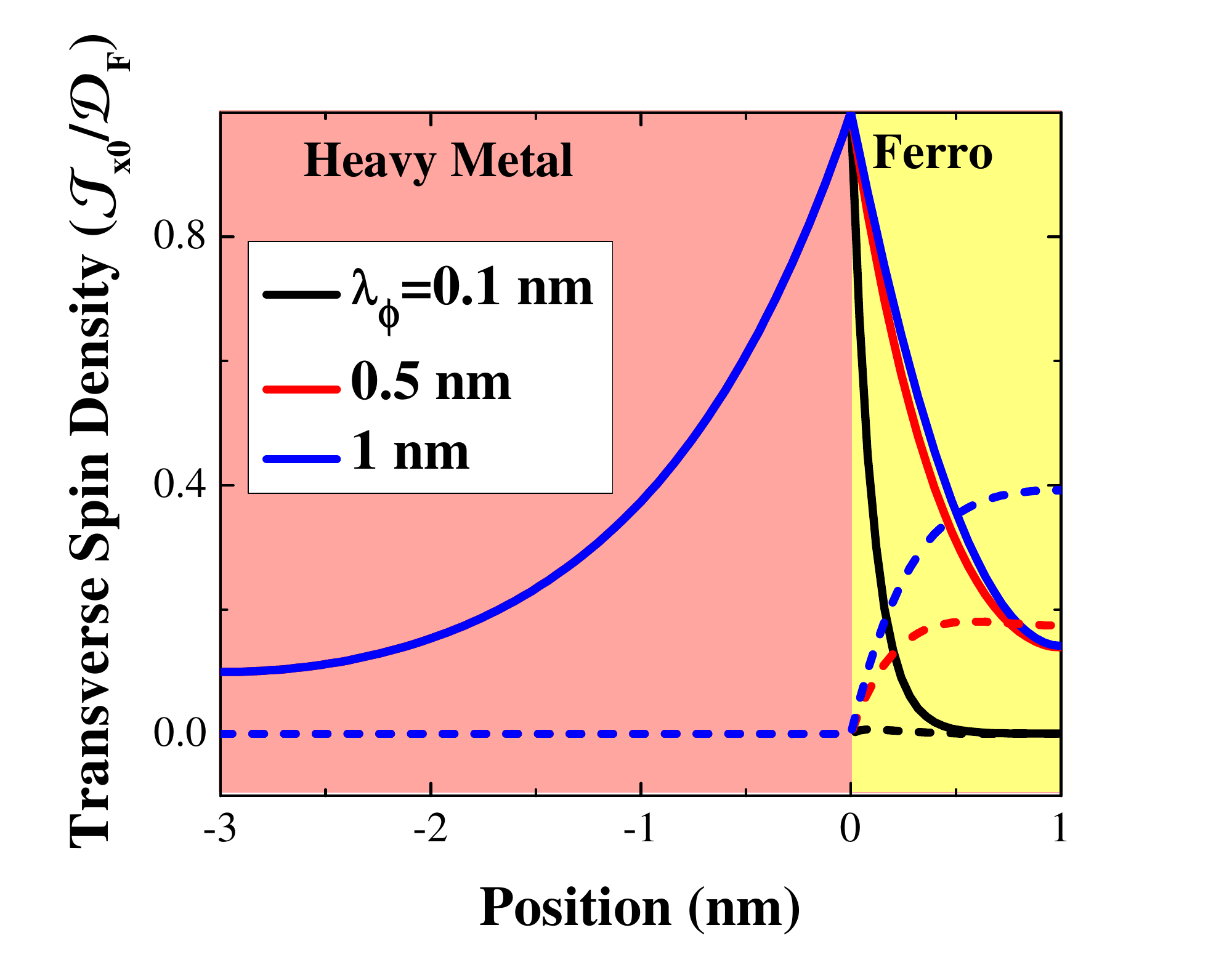}
	\caption{\label{fig:Fig6}(Color online) Case 3: Transverse spin accumulation $S_x$ (solid lines) and $S_y$ (dashed lines) profile along $z$ in the presence of Rashba spin-orbit coupling at the HM/F interface only, for different spin dephasing lengths. The parameters are the same as in Fig. \ref{fig:Fig2}.}
\end{figure}

\begin{figure}
	\centering
		\includegraphics[width=8cm]{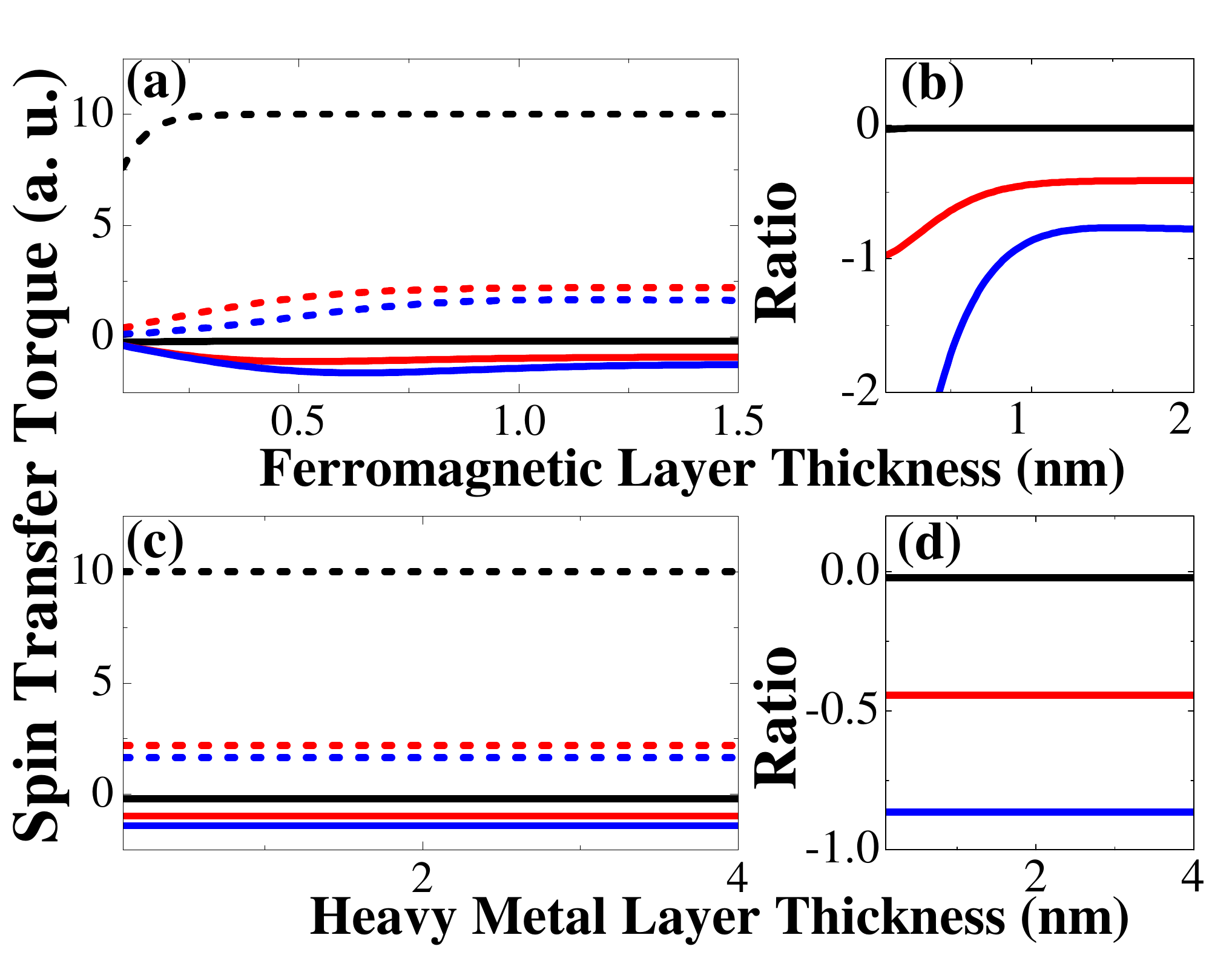}
	\caption{\label{fig:Fig7}(Color online) Case 3: In-plane ($T_{||}$ - solid lines) and out-of-plane torques ($T_\bot$ - dashed lines) as a function of the F (a) and HM (c) layers thickness for different spin dephasing lengths $\lambda_\phi$; spin torque ratio $T_{||}/T_\bot$ as a function of the F (b) and HM (d) layers thickness. The parameters are the same as in Fig. \ref{fig:Fig2}.}
\end{figure}

The Rashba spin-orbit coupling at HM/F interface produces a non-equilibrium spin density along ${\bf S}_0=S^0_x{\bf x}+S^0_y{\bf y}\approx S_x^0{\bf x}$ \cite{wang-manchon-2011} that decays in both F and HM layers. Since the source of the spin accumulation is confined at the HM/F interface, the decay in HM layer is independent on the spin dephasing length (see Fig. \ref{fig:Fig6}).
The thickness dependence of the resulting absolute torque $d {\bf T}$ is represented in Fig. \ref{fig:Fig7}. Again, the HM layer has no influence on the torque magnitude and reducing the F layer thickness can lead to an enhancement of the in-plane torque compared to the perpendicular torque, as shown in Fig. \ref{fig:Fig7}(b). Note however that this variation is only present in layers with thicknesses smaller than the spin dephasing length ($d<\lambda_{\phi}$). Therefore, in realistic systems, no significant difference from the bulk value is expected.

\subsection{Discussion}
From the calculations above, we can draw three main conclusions. First, both Rashba- and SHE-induced torques are on the form ${\bf T}=T_{||}{\bf m}\times{\bf y}+T_\bot{\bf m}\times({\bf y}\times{\bf m})$, the relative magnitude of $T_{||}$ and $T_\bot$ is strongly dependent on the microscopic mechanism such as spin precession, dephasing and relaxation in the bilayer. Second, due to this complex spin dynamics, the ratio $T_{||}/T_\bot$ strongly depends on the thickness of the layers. Third, while the three cases are affected by the ferromagnetic layer thickness, only the SHE-induced torque is affected by the thickness of the heavy metal. This influence is constrained to thicknesses smaller than the spin diffusion length of the HM layer $\lambda_{sf}^{HM}$ since the torque magnitude saturates beyond this length.\par

Note that other contributions of the thickness dependence have been disregarded at this point and have to be accounted to accurately reproduce experimental values. First, as mentioned above, the torque has been evaluated at constant current density in the HM layer ($j_{HM}$ in case 1) and at the interface ($j_i$ in cases 2 and 3). This current density should be replaced by its expression in Eq. (\ref{eq:jn}) or Eq. (\ref{eq:ji}). In addition, at such ultrathin thicknesses the effective conductivity of the layer depends on the ratio $\lambda_e/d$, as shown in Fig. \ref{fig:Fig0}. Finally, changing the thickness of the layers probably modifies the properties of the HM/F interface, such as interfacial resistivity and magnetic anisotropy, complexifying the analysis.\par

However, although analyzing the effect of thicknesses variation on the spin torque requires a good understanding of the materials growth and its magnetic implications, the present study indicates that varying the HM layer thickness over a small range (on the order of $\lambda_{sf}^{HM}$) is sufficient to identify the physical origin of the spin-orbit-induced torque.
\section{Conclusion}
Using a semi-classical drift-diffusion description of the spin transport in a bilayer, both Rashba torque and SHE torque have been considered. The spin dynamics in such ultrathin layers has been investigated and the nature of the spin torque in such systems has been identified. We showed that (i) both torques are on the form ${\bf T}=T_{||}{\bf m}\times{\bf y}+T_\bot{\bf m}\times({\bf y}\times{\bf m})$, (ii) the ratio $T_{||}/T_\bot$ strongly depends on the thickness of the layers and (iii) the thickness dependence of the spin torque may provide an indication of the origin of the torque (Rashba- or SHE-induced).
\section*{Acknowledgement}
The author gratefully acknowledges inspiring discussions with M.D. Stiles, K.-J. Lee and T. Valet.

\end{document}